\documentclass[a4paper,11pt,twoside]{article}
\usepackage{graphicx,anst2}

\begin{document}
\makeatletter
\renewcommand{\theequation}{\thesection.\arabic{equation}}
\@addtoreset{equation}{section}
\makeatother
\def\sltwo{{\sl sl}(2)}
\title{\sc\huge Classification of Type A\\ ${\cal N}$-fold 
Supersymmetry}
\author{\Large Hideaki Aoyama$^{\dagger 1}$,
Noriko Nakayama$^{\star 2}$,
Masatoshi Sato$^{\ddagger 3}$\\[7pt]
\Large and Toshiaki Tanaka$^{\dagger 4}$\\[10pt]
\sc $^\dagger$Faculty of Integrated Human Studies\\[3pt]
\sc Kyoto University, Kyoto 606-8501, Japan\\[7pt]
\sc $^\star$Graduate School of Human and
Environmental Studies\\[3pt]
\sc Kyoto University, Kyoto 606-8501, Japan\\[7pt]
\sc $^\ddagger$The Institute for Solid State Physics\\[3pt]
\sc The University of Tokyo, Kashiwanoha 5-1-5, \\[3pt]
\sc Kashiwa-shi, Chiba 277-8581, Japan}
\footnotetext[1]{aoyama@phys.h.kyoto-u.ac.jp}
\footnotetext[2]{nakayama@phys.h.kyoto-u.ac.jp}
\footnotetext[3]{msato@issp.u-tokyo.ac.jp}
\footnotetext[4]{ttanaka@phys.h.kyoto-u.ac.jp}
\maketitle

\vspace{-12cm}
\rightline{KUCP-0188}
\vspace{11.5cm}
\thispagestyle{empty}
\def\nfsusy{${\cal N}$-fold supersymmetry}
\def\nfsusic{${\cal N}$-fold supersymmetric}
\def\hn{H^-_{\cal N}}
\def\thn{\widetilde H^-_{\cal N}}
\def\hnp{H^+_{\cal N}}
\def\ddq{\frac{d}{dq}}
\def\ww{\widetilde{W}}
\def\wwq{\widetilde{W}(q)}
\def\wD{\widetilde{D}}
\def\ca{\cosh(\alpha q)}
\def\can#1{\cosh^{#1}(\alpha q)}
\def\si{\sinh(\alpha q)}
\def\sian#1{\sinh^{#1}(\alpha q)}
\def\tgn#1{\tanh^{#1}(\gamma q)}
\def\app{a_{++}}
\def\azz{a_{00}}
\def\apz{a_{+0}}
\def\bbp{\bar{b}_+}
\def\bbz{b_0}
\def\bbm{\bar{b}_-}
\def\ddq{\frac{d}{dq}}
\def\cob{\cosh(\beta\, q)}
\def\cobn#1{\cosh^{#1}(\beta\, q)}

\begin{abstract}
Type A ${\cal N}$-fold supersymmetry of one-dimensional quantum
mechanics can be constructed by using {\sl sl}(2)
generators represented on a finite dimensional functional space.
Using this {\sl sl}(2) formalism we show a general method of
constructing Type A ${\cal N}$-fold supersymmetric models.
We also present systematic generation of known models
and several new models using this method.
\end{abstract}

\newpage

\section{Introduction}
\nfsusy\ is an extension of the ordinary supersymmetry in 
one-dimensional quantum mechanics and in fact reduces to it
for ${\cal N}=1$ \cite{AKOSW}--\cite{ANST}.
It has supercharges that are ${\cal N}$-th polynomials of momentum.
(Similar higher derivative generalizations of supercharges were also
investigated in different contexts \cite{AIS}-\cite{FNN}.)
It shares the nonrenormalization theorems of ordinary supersymmetry
\cite{Wit,Wit2}: Some of the energy eigenvalues can be obtained 
in a closed form when \nfsusy\ is not spontaneously broken, while
perturbative parts of some energy eigenvalues are obtained otherwise.
This property is related to what has been known as ``quasi-solvability"
\cite{Tur}--\cite{KP2}.

Recently we have shown that the Type A variety of \nfsusy\
can be constructed using \sltwo\ generators, which are 
defined on a functional space formed by solvable energy
eigenstates \cite{ANST}.
Using this method of construction, we will in this paper 
classify models of Type A \nfsusy, some of which are new.

In the following, we first give a brief review of \nfsusy\
and its construction by the use of \sltwo\ in Section 2.
In Section 3, we first explain the method of model construction and
the way the models are classified, using the fact that
physical content of the model remains invariant under
a linear transformation of a function $h$ that is used to form
a basis of the quasi-solvable space.
Then in the rest of Section 3 we construct a representative
model for each class.  Section 4 contains summary and discussions.

\section{Type A \nfsusy\ and \sltwo}
We will first briefly review the definitions and the fundamental properties
of \nfsusy, its Type A subclass, and the relation with \sltwo \cite{ANST}.

An \nfsusy\ model in one-dimensional quantum mechanics has a Hamiltonian
${\bf H}_{\cal N}$ of the form;
\begin{eqnarray}
{\bf H}_{\cal N}=H^{-}_{{\cal N}}\psi\psi^{\dagger}
+H^{+}_{{\cal N}}\psi^{\dagger}\psi,
\end{eqnarray}
where $\psi$ and $\psi^\dagger$ are fermionic coordinates;
\begin{eqnarray}
\{\psi,\psi\}=\{\psi^{\dagger}, \psi^{\dagger}\}=0
,\quad 
\{\psi,\psi^{\dagger}\}=1,
\label{eq:ferpro}
\end{eqnarray}
and $H^{\pm}_{\cal N}$ are ordinary Hamiltonians for 
one ordinary (bosonic) coordinate $q$;
\begin{eqnarray}
H^{\pm}_{\cal N}=\frac{1}{2}p^{2}+V^{\pm}_{\cal N}(q)
\label{eq:Hfutu}
\end{eqnarray}
with $p=-id/dq$.

The ${\cal N}$-fold supercharges are generically defined as 
\begin{eqnarray}
Q_{\cal N}=P^{\dagger}_{\cal N}\psi
,\quad
Q^{\dagger}_{\cal N}=P_{\cal N}\psi^{\dagger}, 
\end{eqnarray}
where $P_{\cal N}$ is an ${\cal N}$-th order polynomial of $p$:
\begin{eqnarray}
P_{\cal N}=
w_{\cal N}(q)\,p^{\cal N}+w_{{\cal N}-1}(q)\,p^{{\cal N}-1}+\cdots+
w_1(q)\,p+w_0(q). 
\label{sec_dons:pn}
\end{eqnarray}
These satisfy the following \nfsusy\ algebra;
\begin{eqnarray}
&&\{Q_{\cal N}, Q_{\cal N}\} =
\{Q^{\dagger}_{\cal N}, Q^{\dagger}_{\cal N}\}=0,
\label{eq:nilp}\\
&&[\,Q_{\cal N}, {\bf H}_{\cal N}\,] = 
[\,Q^{\dagger}_{\cal N}, {\bf H}_{\cal N}\,]=0.
\label{eq:secondnfsusy}
\end{eqnarray}
In addition, the anticommutator $\{Q_{\cal N}, Q_{\cal N}^\dagger\}$
induces ``mother Hamiltonian", whose relation to the Hamiltonian 
(\ref{eq:Hfutu}) is discussed in detail in Ref.\cite{AST2}.
The nilpotency (\ref{eq:nilp}) is trivially satisfied
due to the property of the fermionic coordinates (\ref{eq:ferpro}).
And the latter relation (\ref{eq:secondnfsusy}) leads to 
\begin{equation}
P_{\cal N}\hn-H^+_{\cal N}P_{\cal N}=0,
\label{eq:nfsusyalgebra}
\end{equation}
and its conjugate.
The identity (\ref{eq:nfsusyalgebra}) induces differential equations
for the functions $V^\pm_{\cal N}(q)$ and $w_n(q)$ $(n=0,1,\cdots,{\cal N})$,
which cannot be solved in general.

In Ref.\cite{AST2} we showed that when a quasi-solvability condition
is satisfied an \nfsusic\ model can be constructed.
In Ref.\cite{ANST} we used this method of construction for the Type A 
subclass of \nfsusy, which is defined by the following form
of the supercharges \cite{AST}:
\begin{eqnarray}
P_{\cal N}=
\left(\wD+i \, \frac{\,{\cal N}-1}{2}E(q)\right)
\left(\wD+i\,\frac{\,{\cal N}-3}{2}E(q)\right)
\cdots
\left(\wD-i\,\frac{\,{\cal N}-1}{2}E(q)\right)_{\,,}
\label{eq:typeapn}
\end{eqnarray}
where $\wD\equiv p-i\wwq$.
We choose the functional subspace where solvable energy eigenstates
belong as the space ${\cal V}$ spanned by the bases 
$\{1, h(q), h(q)^2, \cdots, h(q)^{{\cal N}-1}\}U^{-1}$,
where $h(q)$ is a solution of a differential equation
\begin{equation}
E(q)=\frac{h(q)''}{h(q)'}_{\,,}
\label{eq:ehrel}
\end{equation}
(prime denotes derivative with respect to $q$) and 
\begin{equation}
U\equiv\exp\left[\int\left(\wwq+\frac{\,{\cal N}-1}{2}E(q)\right)dq\right]_{\,.}
\end{equation}
We showed that the Hamiltonian $\hn$ of Type A \nfsusy\ 
is always written as;
\begin{equation}
\hn=U^{-1}\Biggl[-
\begin{array}[t]{c}
\displaystyle\sum_{i,j=+,0,-}\\[-3pt]
\scriptstyle j\ge i
\end{array} 
a_{ij}J^iJ^j + \sum_{i=+,0,-}b_iJ^i + C \,
\Biggr]\,U,
\label{eq:sl2desu}
\end{equation}
where $J^i$ are \sltwo\ generators on $\{1, h, h^2, \cdots, h^{{\cal N}-1}\}$;
\begin{eqnarray}
J^+ \equiv h^2\frac{d}{dh}-({\cal N}-1)h, \quad
J^0 \equiv h\frac{d}{dh}-\frac{{\cal N}-1}{2}, \quad
J^- \equiv \frac{d}{dh},
\label{eq:first}
\end{eqnarray}
and $a_{ij}, b_i$ and $C$ are constants.
These generators satisfy the algebra,
\begin{equation}
[J^+, J^-]=-2J^0, \quad [J^\pm, J^0\,]=\mp J^\pm,
\end{equation}
and form the following Casimir operator:
\begin{equation}
\frac12\left(J^+J^-+J^-J^+\right)-(J^0)^2=-\frac14({\cal N}^2-1).
\label{eq:Casimir}
\end{equation}
By using this identity,
we choose $a_{+-}=0$ in the Hamiltonian (\ref{eq:sl2desu}) without loosing
generality. As a result, there are eight independent real
parameters $\{a_{++}, a_{+0}, a_{00}, a_{0-}, a_{--}, b_+,$ $b_0,$ $b_-\}$
in the Hamiltonian.

Given the above form of the Hamiltonian, the energy eigenstates are
obtained in a trivial manner:
One can use ${\cal N}\times{\cal N}$ matrix representations of 
the generators $J^i$ and write $U\hn U^{-1}$ as an
${\cal N}\times{\cal N}$ matrix and simply diagonalize it.

In Ref.\cite{ANST}, 
we derived the following conditions
by requiring that the Hamiltonian $\hn$ reduce to 
the canonical form (\ref{eq:Hfutu});
\begin{eqnarray}
P_4(h)&\equiv&a_{++}h^4 + a_{+0}h^3 +a_{00}h^2 + a_{0-}h + a_{--}
\nonumber\\
&=&\frac12(h')^2,
\label{eq:p4aaaa}
\end{eqnarray}
and
\begin{eqnarray}
P_3(h)&\equiv&2({\cal N}-2)a_{++}h^3
+\left(\frac{3{\cal N}-5}{2}a_{+0}+b_+\right)h^2
\nonumber\\&&\mbox{}
+\left(({\cal N}-2)a_{00}+b_0\right) h
+\frac{{\cal N}-1}{2}a_{0-}+b_-
\nonumber\\
&=&h'\left(\ww+\frac{\,{\cal N}-2}{2}E\right)
\label{eq:p3aaaa}
\end{eqnarray}
have to be satisfied.
From the fact that $P_4(h)$ is (at most) a fourth order polynomial of $h$
and by the use of the relation (\ref{eq:ehrel}), 
we found the following condition on $E(q)$:
\begin{equation}
\left(\ddq-2E\right)\left(\ddq-E\right)\ddq\left(\ddq+E\right)E=0,
\label{eq:t1}
\end{equation}
for ${\cal N}\ge 2$.
Also, by comparing the coefficients of the highest order terms of
$P_4(h)$ and $P_3(h)$, we found
\begin{equation}
\left(\ddq-E\right)\ddq\left(\ddq+E\right)\ww=0,
\label{eq:t2}
\end{equation}
for ${\cal N}\ge 3$.
When these conditions are met, 
the potentials $V_{\cal N}^\pm(q)$ are given by the following:
\begin{equation}
V^{\pm}_{\cal N}=\frac{1}{2}\,\ww^2+\frac{\,{\cal N}^2-1}{24}
\left(E^2-2E'\right)\pm\frac{\,{\cal N}}{2}\ww'.
\label{eq:t3}
\end{equation}
It is also possible to derive the conditions 
(\ref{eq:t1})--(\ref{eq:t3}) by a direct calculation of the
\nfsusy\ algebra (\ref{eq:nfsusyalgebra}) \cite{future}.

\section{Model construction}
From the results reviewed in the previous section,
we see that one method of constructing a Type A \nfsusy\ model
is to find a solution $E(q)$ of the differential equation (\ref{eq:t1}),
solve Eq.(\ref{eq:t2}) to find a solution $\wwq$, and
then obtain the potentials $V_{\cal N}^\pm(q)$ 
according to the relation (\ref{eq:t3}).
We analyzed several models in this manner in Ref.\cite{ANST}.

This method, however, is difficult to carry out, or at least cumbersome.
The \sltwo\ construction reviewed above provides a rather simple and compact 
alternative to the above program:
Instead of solving the nonlinear differential equation (\ref{eq:t1}),
we can use its origin, Eq.(\ref{eq:p4aaaa}), which can be solved as,
\begin{equation}
q=\pm\int\frac{dh}{\sqrt{2P_4(h)}}_{\,.}
\label{eq:qhin}
\end{equation}
Once $h(q)$ is obtained by inverting Eq.(\ref{eq:qhin}), 
we can obtain $E(q)$ according to Eq.(\ref{eq:ehrel}) and
$\wwq$ according to the relation (\ref{eq:p3aaaa}), or
\begin{eqnarray}
\wwq&=&\frac{P_3(h)}{h'}-\frac{\,{\cal N}-2}{2}E\nonumber\\
&=&\frac1{h'}\left(
P_3(h)-\frac{{\cal N}-2}{2}\frac{d}{dh}P_4(h)
\right)\nonumber\\
&=&\frac1{h'}
\left(\bbp h^2+\bbz h+\bbm \right),
\end{eqnarray}
where we have used a relation 
\begin{equation}
E(q)=\frac{h''}{h'}=\frac1{h'^2}\frac{d}{dq}\left(\frac12h'^2\right)
=\frac1{h'}\frac{d}{dh}P_4(h),
\end{equation}
and absorbed some of the $a$-coefficients in the redefinition of 
$b$-coefficients:
\begin{equation}
\bar{b}_+\equiv b_+ +\frac12 a_{+0},\quad
\bar{b}_-\equiv b_- + \frac12 a_{0-}.
\end{equation}
Once this is done, the potentials are simply given by the expression 
(\ref{eq:t3}) from $E(q)$ and $\wwq$.
They are given as the following functions of $h$, 
\begin{eqnarray}
V_{\cal N}^{\pm}&=&\frac{1}{16 P_4(h)}
\left(({\cal N}^2-1)\left(\frac{d}{dh}P_4(h)\right)^2+4w(h)^2
\mp 4{\cal N}w(h)\frac{d}{dh}P_4(h)\right) 
\nonumber\\
&&\mbox{}-\frac{{\cal N}^2-1}{12}\frac{d^2}{dh^2}P_{4}(h)
\pm\frac{{\cal N}}{2}\frac{d}{dh}w(h), 
\end{eqnarray}
where $w(h)=\bar{b}_{+}h^2+\bbz h+\bar{b}_{-}$.

Unfortunately, for the most general fourth order polynomial $P_4(h)$, 
the integral in Eq.(\ref{eq:qhin}) involves elliptic integrals, and 
therefore the function $h(q)$ cannot be written down explicitly.
For this reason, we will start with the most simple $P_4(h)$ and proceed to
more complex ones.  
In doing so, it is important to classify resulting models, as many of them are
equivalent to each other.  This is because of the following:
Let us define a new function $\hat{h}$ by the following;
\begin{equation}
h = s \hat{h} + t,
\label{eq:lintra}
\end{equation}
where $s$ and $t$ are $q$-independent constants.
Using $\hat{h}$ in place of $h$ 
does not affect the physical content of the model,
because $h$ is introduced to 
form the basis $\{1, h, h^2, \cdots, h^{{\cal N}-1}\}$
that defines the space ${\cal V}$
and 
the new base $\{1, \hat{h}, \hat{h}^2, \cdots, \hat{h}^{{\cal N}-1}\}$
defines the same space of ${\cal V}$.
This invariance under the linear transformation (\ref{eq:lintra})
is also evident in the expression $E(q)$ in Eq.(\ref{eq:ehrel}).
A caution, however, is needed for $P_4(h)$.
In order for Eq.(\ref{eq:p4aaaa}) to be satisfied with $\hat{h}$,
or equivalently, for $q$ to be invariant 
(up to the trivial integration constant)
in Eq.(\ref{eq:qhin}), we need to replace as $P_4(h)$ as follows:
\begin{equation}
P_4(h) \rightarrow \hat{P}_4(\hat{h})=\frac1{s^2}P_4(h).
\label{eq:p4tra}
\end{equation}
As for $\wwq$, in doing the replacement
(\ref{eq:lintra}) we can transform the coefficients
so that $\wwq$ remains invariant.
This way, the potentials remain invariant under the 
transformation (\ref{eq:lintra}) and (\ref{eq:p4tra}).

It is also important to note
that on the right hand side of Eq.(\ref{eq:qhin}),
even a complex integration constant is allowed, 
since it still is the solution of Eq.(\ref{eq:p4aaaa}).
Therefore, we can freely perform the following complex translation
of $q$ in $h(q)$, $E(q)$, $\wwq$ and 
$V^{\pm}_{\cal N}(q)$;
\begin{eqnarray}
q \rightarrow q+q_0,
\end{eqnarray}
where $q_0$ is an arbitrary complex constant.
We will also use this when necessary.

In the following construction of model, 
we will classify the resulting models according to the distribution of
zero points (and their orders) of $P_4(h)$.
And then we will use the transformation (\ref{eq:lintra}) and (\ref{eq:p4tra})
to choose representatives of each class by
moving the zero points of $P_4(h)$ to convenient locations;
typically $h=0$ and $|h|=1$, with the zero point $h=0$ having the
highest order, as much as we can.
(In doing so, we will avoid 
distinguishing $\hat{h}$ from $h$ and $\hat{P}_4$ from $P_4$
unless it is necessary.)
We will denote the resulting models by the set of the 
zero points of $P_4(h)$ with the order of each zero point as a
superindex (if more that one); for example $\{0^2,1\}$ denotes
a model who has a zero point of order 2 at $h=0$ and a zero point of
order 1 at $h=1$.
Also, in the following we will choose the sign on the right hand side of 
Eq.(\ref{eq:qhin}), by using reflection of $q$,
so that the resulting expression is free from extra minus signs.

\subsection{Case (0): Constant $\mathbf{P_4(h); \{\phi\}}$}
The most simple model is given by a constant $P_4(h)$.
For $P_4(h)=a_{--}$, we obtain the following:
\begin{equation}
h(q)=\sqrt{2a_{--}}q,
\end{equation}
which induces,
\begin{eqnarray}
E(q)&=&0,\\
\wwq &=& \sqrt{2a_{--}}\,\bar{b}_+ q^2 + \bbz q + 
\frac{\bar{b}_-}{\sqrt{2a_{--}}}_{\,.}
\end{eqnarray}

Applying the transformation (\ref{eq:p4tra}), 
we can get rid of the parameter $a_{--}$ (and still not loose the generality).
We choose to do so with $s=\sqrt{2a_{--}}$, so that we have
$P_4(h)=1/2$, which is equivalent to setting $a_{--}=1/2$ to start with.
This way, we obtain the following:
\begin{eqnarray}
h(q)&=&q,\\
E(q)&=&0,\\
\wwq &=& \bar{b}_+ q^2 + \bbz q + \bar{b}_-,\\
V_{\cal N}^\pm (q)&=& \frac{\,\bar{b}_+^2}{2}q^4+\bar{b}_+\bbz\, q^3
+\left(\bar{b}_+\bar{b}_-+\frac12\bbz^2\right)q^2
+\left(\bbz\bar{b}_-\pm{\cal N}\bar{b}_+\right)q
+\frac12\left(\bar{b}_-^2 \pm {\cal N}\bbz\right)_{\,.}\nonumber\\
\end{eqnarray}
With suitable definition of parameters,
the above potentials become asymmetric double-well potentials,
whose \nfsusy\ is spontaneously broken by nonperturbative effects.
The solvable energy eigenvalues represent only the perturbative parts
(by suitable definition of the coupling constant).
Its nonperturbative aspects and the asymptotic behaviors of the
perturbation series were studied 
in Refs.\cite{AKOSW,AKHSW,AKHOSW} and resulted in the discovery of
\nfsusy\ \cite{AKOSW,AKOSW2}.

\subsection{Case (1): Linear $\mathbf{P_4(h): \{0\}}$}
The function $P_4(h)=a_{0-}h+a_{--}$ leads to the following when
the integration constant is appropriately chosen:
\begin{equation}
h(q)=\frac{\,a_{0-}}{2}q^2+\sqrt{2a_{--}}q, 
\end{equation}
which results in;
\begin{eqnarray}
\quad E(q)&=&\frac{a_{0-}}{\displaystyle a_{0-}q+\sqrt{2a_{--}}}_{\,,}\\
\wwq&=&\frac{1}{\displaystyle 4(a_{0-}q+\sqrt{2a_{--}})}
\biggl[a_{0-}^2\bar{b}_+q^4
+4\sqrt{2a_{--}}a_{0-}\bar{b}_+q^3
+(8a_{--}\bar{b}_++2a_{0-}\bbz)q^2\nonumber\\
&&\mbox{}\qquad+4\sqrt{2a_{--}}\,\bbz \,q
+4\bar{b}_-\biggr]_{\,.}
\end{eqnarray}
If we set $a_{0-}=0$, we reproduce Case (0) studied above.

For $a_{0-}\ne0$, the above lead to potentials that are 
rational polynomials of $q$.
In order to obtain a simple representative model, 
we apply the translation ($t$) to move the zero point of 
$P_4(h)$ to $h=0$ and use the scaling ($s$) to set
$P_4(h)=2h$. This leads to the following:
\begin{eqnarray}
h(q)&=&q^2,\\
E(q)&=&\frac1q,\\
\wwq&=&\frac12\left[
\bar{b}_+ q^3+ \bbz q+ \frac{\,\bar{b}_-}{q}\right].
\end{eqnarray}
The potentials are given by the following;
\begin{eqnarray}
V_{\cal N}^\pm (q)&=&
\frac{\,\bar{b}_+^2}{8}\,q^6
+\frac{\,\bar{b}_+ \bbz}{4}\,q^4
+\frac18\left(\bbz^2+2\bar{b}_+\bar{b}_-\pm6{\cal N}\,\bar{b}_+\right)q^2
\nonumber\\
&&\mbox{}+\frac18\left(\bar{b}_-^2+{\cal N}^2-1\mp2{\cal N}\bar{b}_-\right)
\frac1{\,q^2}
+\frac14\bbz\left(\bar{b}_- \pm{\cal N}\right).
\end{eqnarray}
This model contains the sextic potentials studied in Refs.\cite{AST2,DDT,KP2}
as special cases.

\subsection{Case (2): Quadratic $\mathbf{P_4(h)}$}
The general quadratic function $P_4(h)=a_{00}h^2 + a_{0-}h + a_{--}$
leads to the following;
\begin{eqnarray}
h(q)&=&\frac{a_{0-}}{2a_{00}}
\left(\cosh(\alpha q)-1\right)
+\sqrt{\frac{a_{--}}{a_{00}}}\sinh(\alpha q)_{\,,}
\label{eq:case2h}\\
E(q)&=&\alpha\frac
{\,a_{0-}\cosh(\alpha q)
+2\sqrt{a_{--}a_{00}}\sinh(\alpha q)}
{\,a_{0-}\sinh(\alpha q)
+2\sqrt{a_{--}a_{00}}\cosh(\alpha q)}_{\,,}
\end{eqnarray}
where $\alpha\equiv\sqrt{2a_{00}}$.
Reproduction to Case (1) for $a_{00}=0$ is evident in these expressions.

Models in this category are classified to two cases;
(2a) one zero point of order 2,
and 
(2b) two different zero points.
We will choose 
the zero points for each cases as;
(2a) $\{0^2 \}$ and (2b) $\{\pm1\}$.
There is also a notable variation to (2b), namely Case (2b$'$) 
with zero points at $\{\pm i\}$, which will be also studied.

\vskip 4mm \noindent{\bf Case (2a): $\mathbf{\{0^2\}}$}\newline
In this case, we choose $P_4(h)=a_{00}h^2$.
Note that since $a_{00}$ is invariant under the transformation
(\ref{eq:p4tra}), it cannot be set to a particular value without
loosing generality. 
We obtain the following:
\begin{eqnarray}
h(q)&=&e^{\alpha q},
\label{eq:case2h1}\\
E(q)&=&\alpha,\\
\wwq&=&\frac1{\alpha}
\left[\,\bar{b}_+e^{\alpha q}+\bbz+\bar{b}_-e^{-\alpha q}\right],\\
V_{\cal N}^\pm(q) &=&
\frac{\bar{b}_+^2}{4a_{00}}\,e^{2\alpha q}
+\frac{\bar{b}_+}{2} \left(\frac{\bbz}{a_{00}}\pm{\cal N}\right)
e^{\alpha q}
+\frac{\bar{b}_-}{2} \left(\frac{\bbz}{a_{00}}\mp{\cal N}\right)
e^{-\alpha q}
\nonumber\\
&&\mbox{}+
\frac{\bar{b}_-^2}{4a_{00}}e^{-2\alpha q}
+\frac{\bbz^2}{4a_{00}}+\frac{\bar{b}_+\bar{b}_-}{2a_{00}}
+\frac{a_{00}}{12}\left({\cal N}^2-1\right)_{.}
\end{eqnarray}
These induces exponential potentials for $a_{00}>0$ 
and  periodic potentials for $a_{00}<0$, whose special cases
were studied in Refs.\cite{ASTY, KP1}.

\vskip 4mm\noindent{\bf Case (2b): $\mathbf{\{\pm1\}}$}\newline
With $P_4(h)=a_{00}(h^2-1)$, we obtain the following:
\begin{eqnarray}
h(q)&=&\cosh(\alpha q),\\
E(q)&=&\frac{\alpha}{\tanh(\alpha q)}_{\,,}\\
\wwq&=&\frac1{\alpha\sinh(\alpha q)}
\left[\,\bar{b}_+\cosh^2(\alpha q)+ \bbz\cosh(\alpha q)
+\bar{b}_-\right]_{\,,}\\
V_{\cal N}^\pm(q) &=&
\frac{1}{\sinh^2(\alpha q)}
\Biggl[
\frac{\bbp^2}{4\azz}\can4
+\frac12\left(\frac{\bbp\bbz}{\azz} \pm{\cal N}\bbp\right)\can3
\nonumber\\&&\mbox{}
+\left(\frac{\bbz^2+2\bbp\bbm}{4\azz}+\frac{{\cal N}^2-1}{12}\azz\right)\can2
+\left(\frac{\bbz\bbm}{2\azz}
\mp{\cal N}\left(\bbp+\frac{\bbm}{2}\right)\right)\ca
\nonumber\\&&\mbox{}
+\left(\frac{\bbm^2}{4\azz}+\frac{{\cal N}^2-1}{6}\azz
\mp\frac{\cal N}{2}\bbz\right)\Biggr]_{\,.}
\end{eqnarray}

\vskip 4mm \noindent{\bf Case (2b$'$): $\mathbf{\{\pm i\}}$}\newline
An alternative to Case (2b) is $P_4(h)=a_{00}(h^2+1)$.
Although this case is obtained by a scaling with $s=i$ from Case (2b),
it is listed here since the resulting expression may be useful due to the 
lack of poles in the potentials.
In this case, from Eq.(\ref{eq:qhin}) we obtain
\begin{eqnarray}
h(q)&=&\sinh(\alpha q).
\label{eq:case2bp}
\end{eqnarray}
Note that we have chosen the integration constant differently
from Case (2b).
As a result, in order to obtain the above and the rest of the formula
of this case from Case (2b),
one needs to do a complex translation $q\rightarrow q+\pi i/(2\alpha)$
and a scaling $h\rightarrow ih$.
Eq.(\ref{eq:case2bp}) leads to the following:
\begin{eqnarray}
E(q)&=&\alpha \tanh(\alpha q),\\
\wwq&=&
\frac1{\alpha\cosh(\alpha q)}
\left[\,\bar{b}_+\sinh^2(\alpha q)
+\bbz\sinh(\alpha q)+\bar{b}_-\right]_{\,,}\\
V_{\cal N}^\pm(q) &=&
\frac{1}{\can2}
\Biggl[
\frac{\bbp^2}{4\azz}\sian4
+\frac12\left(\frac{\bbp\bbz}{\azz} \pm{\cal N}\bbp\right)\sian3
\nonumber\\&&\mbox{}
+\left(\frac{\bbz^2+2\bbp\bbm}{4\azz}+\frac{{\cal N}^2-1}{12}\azz\right)\sian2
+\left(\frac{\bbz\bbm}{2\azz}
\pm{\cal N}\left(\bbp-\frac{\bbm}{2}\right)\right)\si
\nonumber\\&&\mbox{}
+\left(\frac{\bbm^2}{4\azz}-\frac{{\cal N}^2-1}{6}\azz
\pm\frac{\cal N}{2}\bbz\right)\Biggr]_{\,.}
\end{eqnarray}

\subsection{Case (3): Cubic $\mathbf{P_4(h)}$}
In case $P_4(h)$ is a general cubic polynomial
we can no longer write down an algebraic expression of $h(q)$, 
as the integral in Eq.(\ref{eq:qhin}) yields expressions involving
elliptic functions.
On the other hand, for a case when $P_4(h)=a_{+0} h^3 + a_{00} h^2$, 
we find that we can write an algebraic expression of $h(q)$.
In this case, we obtain;
\begin{equation}
h(q)=-\frac{\,2a_{00}}{\,a_{+0}}
\frac1{\,\cosh(\alpha q)+1}_{\,.}
\label{eq:case3h}
\end{equation}
In the above, the integration constant in Eq.(\ref{eq:qhin}) is
chosen so that the expression of $h(q)$ is most simple.
In order to obtain $h(q)$ of Eq.(\ref{eq:case2h1}) of Case (2a),
one needs to do a translation on $q$;
\begin{equation}
q \rightarrow q - \sqrt{\frac2{a_{00}}}\,
{\rm arctanh}\sqrt{1+\frac{\,a_{+0}}{a_{00}}}_{\,,}
\end{equation}
first and then take $a_{+0}=0$.

From the fact that we have the closed expression (\ref{eq:case3h}) for 
$P_4(h)=a_{+0} h^3 + a_{00} h^2$, we see that
calculable cases are limited to the cases that have a zero point of 
at least order two.
In the following, we choose the representative models as ones with
zero points being (3a) $\{0^3\}$ and (3b) $\{0^2, 1\}$.

\vskip 4mm \noindent{\bf Case (3a): $\mathbf{\{0^3\}}$}\newline
In this case, we use only the translation to move the
zero point to $h=0$. Using the remaining scaling degree of freedom, 
we can choose
the coefficient of $h^3$ to any value.
We choose $P_4(h)=2h^3$ to obtain the following:
\begin{eqnarray}
h(q)&=&\frac1{\,q^2}_{\,,}\\
E(q)&=&-\frac3{q}_{\,,}\\
\wwq&=&-\frac12\left(\bar{b}_-q^3 
+ \bbz\, q +\frac{\,\bar{b}_+}{q}\right)_{\,,}
\\
V_{\cal N}^\pm(q) &=&
\frac{\,\bar{b}_-^2}{8}q^6
+\frac{\,\bbz\bar{b}_-}{4}q^4
+\frac18
\left(\bbz^2 + 2\bar{b}_+\bar{b}_- \mp6{\cal N}\,\bar{b}_-\right)q^2\nonumber\\
&&\mbox{}
+\frac1{8}
\left(\bar{b}_+^2+{\cal N}^2 -1 \pm2{\cal N}\,\bar{b}_+\right)
\frac1{\,q^2}
+\frac1{4}\bbz\left(\bar{b}_+\mp{\cal N}\right).
\end{eqnarray}

\vskip 4mm \noindent{\bf Case (3b): $\mathbf{\{0^2,1\}}$}\newline
In case we have one zero point of order 2,
we use the translation on $h$ to move it to $h=0$
and use scaling on $h$ to move the other zero point to
$h=1$. In this manner we obtain $P_4(h)=-a_{00}h^2(h-1)$
with $a_{00}$ remaining as an arbitrary parameter.
In this case, we find the following using $\beta\equiv\sqrt{a_{00}/2\,}$\,:
\begin{eqnarray}
h(q)&=&\frac1{\cobn2}_{\,,}\\
E(q)&=&-2\beta\frac{\cosh(2\beta q)-2}{\sinh(2\beta q)}_{\,,}\\
\wwq&=-&\frac1{2\beta\sinh(\beta q)}
\left[\frac{\bar{b}_+}{\cob}+\bbz\cob+\bbm\cobn3\right]_{\,,}\\
V_{\cal N}^\pm(q) &=&
\frac1{\cobn2\sinh^2(\beta \,q)}
\Biggl[
\frac{\bbm^2}{4\azz}\cobn8
+\bbm\left(\frac{\bbz}{2\azz}\mp\frac{\cal N}{2}\right)\cobn6
\nonumber\\&&\mbox{}
+\left(
\frac{\bbz^2+2\bbp\bbm}{4\azz}
+\frac{{\cal N}^2-1}{12}\azz \pm\frac{3\cal N}{4}\bbm
\right)\cobn4
\nonumber\\&&\mbox{}
+\left(
\frac{\bbp\bbz}{2\azz}-\frac{{\cal N}^2-1}{12}\azz
\pm\frac{\cal N}{4}(2\bbp+\bbz)
\right)\cobn2
\nonumber\\&&\mbox{}
+\left(
\frac{\bbp^2}{4\azz}+\frac{{\cal N}^2-1}{16}\azz
\mp\frac{\cal N}{4}\bbp
\right)
\Biggr]_{\,.}
\end{eqnarray}
For $a_{00}>0$, this corresponds to 
the ``Hyperbolic case" noted in Ref.\cite{ANST}.
As in Case (2b$'$), we could make an alternative choice 
$\{0^2,i\}$ of zero points, which, however, leads to
a model that is identical to this case.

\subsection{Case (4): Quartic $\mathbf{P_4(h)}$}
As is already evident in Case (3), we cannot do the integration
in Eq.(\ref{eq:qhin}) for the most general case.
This case, however, is rather important and interesting,
since no models with $a_{++}\ne0$ were listed in Ref.\cite{Tur}
or were constructed as an \nfsusic\ model in the past.
The latter is partly because of the fact that in our
previous construction of \nfsusic\ models \cite{AST2}, reduction on 
${\cal N}$ was used with implicit assumption, which were
satisfied only for $a_{++}=0$, as was explained in detail in Ref.\cite{ANST}.

We find that for $P_4(h)=a_{++}h^4+a_{+0}h^3+a_{00}h^2$
we can obtain expression of $h(q)$ explicitly as follows;
\begin{equation}
h(q)=\frac{\displaystyle 4a_{00}^{3/2}e^{\alpha q}}
{\displaystyle a_{00}-2a_{+0}\sqrt{a_{00}}\,e^{\alpha q}
+(a_{+0}^2-4a_{++}a_{00})\,e^{2\alpha q}}_{\,.}
\label{eq:case4}
\end{equation}
Therefore, we see that closed algebraic expressions can
be written down when at least one zero point is of order 2 or higher.
Such cases can be classified as follows;
(4a) one zero point of order four, with representative model having
$\{0^4\}$,
(4b) one zero point of order 3 and another different zero point,
with $\{0^3,1\}$,
(4c) two zero points, both of them being of order 2, with $\{\pm 1^2\}$,
(4c$'$) variation of (4c) with $\{\pm i^2\}$,
(4d) one zero point of order 2 and two different zero points,
whose representative model is, for example, $\{0^2, 1, \sigma(\ne 1)\}$,
Since Case (4d) results in expressions that are 
not much simpler than Eq.(\ref{eq:case4}), we will
skip it and study its special cases of
(4d1) $\{0^2, \pm 1\}$, and 
(4d1$'$) $\{0^2, \pm i\}$, which are related to each other by
complex scaling on $h$.

\vskip 4mm \noindent{\bf Case (4a): $\mathbf{\{0^4\}}$}
\newline
In this case, we choose $P_4(h)=h^4/2$, 
since only the translation is used to move the zero point of 
order 4 to $h=0$ and we have the freedom to choose the scaling on $h$.
This choice leads to the following:
\begin{eqnarray}
h(q)&=& \frac1q_{\,,}\\
E(q) &=& -\frac2q_{\,,}\\
\wwq &=& -\bbm q^2 -\bbz q -\bbp,\\
V_{\cal N}^\pm(q) &=& 
\frac{\,\bbm^2}{2}q^4
+\bbz\bbm q^3
+\frac12\left(\bbz^2+2\bbp\bbm\right)q^2
+\left(\bbp\bbz \mp{\cal N}\bbm\right)q
+\frac12\left(\bbp^2 \mp {\cal N}\bbz\right)_{\,.}\nonumber\\
\end{eqnarray}

\vskip 4mm \noindent{\bf Case (4b): $\mathbf{\{0^3, 1\}}$}
\newline
For $P_4(h)=a_{++}h^3(h-1)$, we find the following:
\begin{eqnarray}
h(q)&=&-\frac{2}{\app q^2-2}_{\,,}\\
E(q) &=&-\frac{3\app q^2 +2}{q(\app q^2-2)}_{\,,}\\
\wwq &=&\frac{\,\app\bbm}{4}q^3
-\left(\bbm+\frac{\,\bbz}{2}\right)q
+\frac{\bbp+\bbz+\bbm}{\app}\frac1q_{\,,}\\
V_{\cal N}^\pm(q) &=&
\frac{\,\app^2\bbm^2}{32}q^6
-\frac{\,\app\bbm(2\bbm+\bbz)}{8}q^4
+\frac18\left(\bbm(2\bbp+6\bbz+6\bbm) +\bbz^2\pm3{\cal N}\app\bbm\right)q^2
\nonumber\\&&\mbox{}
+\left(\frac{(\bbp+\bbz+\bbm)^2}{2\app^2}
+\frac{{\cal N}^2-1}{8}\mp\frac{\cal N}{2\app}\left(\bbp+\bbz+\bbm\right)
\right)\frac1{q^2}
\nonumber\\&&\mbox{}
-\frac{(2\bbm+\bbz)(2(\bbp+\bbz+\bbm)\pm{\cal N}\app)}{4\app}_{\,.}
\end{eqnarray}

\vskip 4mm \noindent{\bf Case (4c): $\mathbf{\{\pm1^2\}}$}
\newline
For $P_4(h)=a_{++}(h^2-1)^2$, we find the following 
with $\displaystyle\gamma\equiv\sqrt{2\app}$:
\begin{eqnarray}
h(q)&=&\frac1{\tanh(\gamma q)}_{\,,}\\
E(q) &=&-\frac{2\gamma}{\tanh(\gamma q)}_{\,,}\\
\wwq &=&-\frac{1}{\gamma}\sinh^2(\gamma q)
\left[\frac{\bbp}{\tgn2}+\frac{\bbz}{\tgn{}}+\bbm\right]_{\,.}
\end{eqnarray}
The expression of the potentials $V_{\cal N}^\pm(q)$ is rather long
and not particularly informative and therefore we will not list it here.

Similarly to Case (2b$'$), one could move the zero points of this case to
$h=\pm i$.  This results in $h(q)=\tan(\gamma q)$, 
$E(q)=2\gamma\tan(\gamma q)$.

\vskip 4mm \noindent{\bf Case (4d1): $\mathbf{\{0^2, \pm1\}}$}
\newline
For $P_4(h)=-a_{00}h^2(h^2-1)$, we find the following:
\begin{eqnarray}
h(q)&=& \frac1{\cosh(\alpha q)}_{\,,}\\
E(q) &=& \frac{\alpha}{\sinh(\alpha q)\cosh(\alpha q)}
\left[\, 2 - \cosh^2(\alpha q)\right]_{\,,}\\
\wwq &=& -\frac1{\alpha\sinh(\alpha q)}
\left[\bar{b}_++\bbz \cosh(\alpha q)+\bar{b}_- \cosh^2(\alpha q)
\right]_{\,,}\\
V_{\cal N}^\pm(q) &=&
\frac1{\sinh^2(\alpha \,q)}
\Biggl[
\frac{\bbm^2}{4\azz}\can4
+\frac{\bbm}{2}\left(\frac{\bbz}{\azz}\mp{\cal N}\right)\can3
\nonumber\\&&\mbox{}
+\left(\frac{\bbz^2+2\bbp\bbm}{4\azz}+\frac{{\cal N}^2-1}{12}\azz\right)\can2
+\left(\frac{\bbp\bbz}{2\azz}\pm\frac{\cal N}{2}(\bbp+2\bbm)\right)\can{}
\nonumber\\&&\mbox{}
+\left(\frac{\bbp^2}{4\azz}+\frac{{\cal N}^2-1}{6}\azz
\pm\frac{\cal N}{2}\bbz\right)\Biggr]_{\,.}
\end{eqnarray}

\vskip 4mm \noindent{\bf Case (4d1$'$): $\mathbf{\{0^2, \pm i\}}$}\newline
For $P_4(h)=a_{00}h^2(h^2+1)$, we find the following:
\begin{eqnarray}
h(q)&=& \frac1{\sinh(\alpha q)},\\
E(q) &=& \frac{\alpha}{\sinh(\alpha q)\cosh(\alpha q)}
\left[\, -2 - \sinh^2(\alpha q)\right].
\end{eqnarray}
As noted above, this case is obtained from Case (4d1) 
by $h\rightarrow ih$ and therefore the rest of the
expressions are not listed here.

\section{Summary and Discussions}
In this paper, we have classified the Type A \nfsusy\ models
according to the zero point structure of $P_4(h)$, which is at most
a fourth order polynomial of $h$.  
We have found that the potentials of the model can
be written down explicitly for the most general distribution of 
the zero points only when the function $P_4(h)$
is at most second order polynomial of $h$.
In such cases, we have classified the resulting models
and have written down the potentials for representative cases of each classes.
When the function $P_4(h)$ is a third or forth order polynomial of $h$,
we have found that the potentials can be written down
explicitly only when at least one zero point is of order 2 or higher.
For those cases, we have written down relevant expressions for
each representative cases.
Along the way, we have found several new models not studied previously.

It should be stressed that all these models are 
connected to each other in the eight-parameter space
$\{a_{++}, a_{+0}, a_{00}, a_{0-}, a_{--}, b_+,$ $b_0,$ $b_-\}$.
Evidently, many of them are worth studying in detail, because
of the quasi-solvability, as is clear in the expression of
the Hamiltonian (\ref{eq:sl2desu}). 
We also note that as far as the resulting 
potentials are concerned, some of the cases listed in this
paper are equivalent.  For example, potentials $V_{\cal N}^\pm(q)$
of Case (0) and $V_{\cal N}^\mp(q)$ of Case (4a)
are equivalent to each other when 
coefficients $\bbp$ and $\bbm$ are interchanged.
Similarly, Cases (1), (3a) and (4b), Cases (2a) and (4c), and, Cases
(2b), (3b) and (4d1) have equivalent potentials.
They, however, are listed here as separate cases, since 
they all differ in the form of supercharges.
This equivalence/difference issue may be worth studying further.

\section*{Acknowledgments}
H. Aoyama's work was supported in part by the Grant-in-Aid for
Scientific Research No.10640259.
T. Tanaka's work was supported in part by a JSPS research fellowship.

\section*{Note added in proof}

After the completion of this work, we came across an interesting 
 paper, A. Gonz\'alez-L\'opez, N. Kamran, and P. J. Olver, {\sl Commun. 
Math. Phys.} {\bf 153} (1993) 117,  where $GL(2,\mathbf{R})$ symmetry in quasi-solvable 
systems is discussed.  The equivalence of the potentials mentioned 
 in the summary may be explained naturally by this symmetry.

\end{document}